\documentclass[superscriptaddress,amsmath,amssymb,aps,prl,twocolumn,floatfix]{revtex4}

\usepackage{graphicx}
\usepackage{dcolumn}
\usepackage{bm}
\usepackage[colorlinks=true,urlcolor=magenta,linkcolor=cyan,citecolor=magenta]{hyperref}
\usepackage{lipsum}
\usepackage[utf8]{inputenc}
\usepackage{siunitx}
\usepackage{ulem}

\begin{document}

\title{Twist-dependent intra- and interlayer excitons in moir\'e MoSe$_2$ homobilayers}

\author{Viviana Villafañe}
\thanks{Contributed equally to this work.}
\author{Malte Kremser}
\thanks{Contributed equally to this work.}
\affiliation{Walter Schottky Institut and Physik Department, Technische Universit{\" a}t M{\" u}nchen, Am Coulombwall 4, 85748 Garching, Germany}
\author{Ruven Hübner}
\affiliation{Institut für Theoretische Physik, Universität Bremen, P.O. Box 330 440, 28334 Bremen, Germany}
\author{Marko M. Petri\'{c}}
\affiliation{Walter Schottky Institut and Department of Electrical and Computer Engineering, Technische Universit\"{a}t M\"{u}nchen, Am Coulombwall 4, 85748 Garching, Germany}
\author{Nathan P. Wilson}
\author{Andreas V. Stier}
\affiliation{Walter Schottky Institut and Physik Department, Technische Universit{\" a}t M{\" u}nchen, Am Coulombwall 4, 85748 Garching, Germany}
\author{Kai M\"uller}
\affiliation{Walter Schottky Institut and Department of Electrical and Computer Engineering, Technische Universit\"{a}t M\"{u}nchen, Am Coulombwall 4, 85748 Garching, Germany}
\author{Matthias Florian}
\affiliation{University of Michigan, Dept. of Electrical Engineering and Computer Science, 48109 Ann Arbor, MI, USA}
\author{Alexander Steinhoff}
\affiliation{Institut für Theoretische Physik, Universität Bremen, P.O. Box 330 440, 28334 Bremen, Germany}
\author{Jonathan J. Finley}
\email{finley@wsi.tum.de}
\affiliation{Walter Schottky Institut and Physik Department, Technische Universit{\" a}t M{\" u}nchen, Am Coulombwall 4, 85748 Garching, Germany}

\begin{abstract}
Optoelectronic properties of van der Waals homostructures can be selectively engineered by the relative twist angle between layers. Here, we study the twist-dependent moiré coupling in MoSe$_2$ homobilayers. For small angles, we find a pronounced redshift of the \textbf{K}-\textbf{K} and \boldsymbol{$\Gamma$}-\textbf{K} excitons accompanied by a transition from \textbf{K}-\textbf{K} to \boldsymbol{$\Gamma$}-\textbf{K} emission.
Both effects can be traced back to the underlying moiré pattern in the MoSe$_2$ homobilayers, as confirmed by our low-energy continuum model for different moiré excitons.
We identify two distinct intralayer moiré excitons for R-stacking, while H-stacking yields two degenerate intralayer excitons due to inversion symmetry. In both cases, bright interlayer excitons are found at higher energies.
The performed calculations are in excellent agreement with experiment and allow us to characterize the observed exciton resonances, providing insight about the layer composition and relevant stacking configuration of different moiré exciton species.
\end{abstract}

\maketitle

Van der Waals (vdW) homo- and heterostructures formed from monolayer transition metal dichalcogenides (TMDs) are unique semiconductor systems in which light couples to electronic and spin excitations with potential for novel optoelectronic and valleytronic applications \cite{Gei13, Liu16}. In angle-aligned TMD heterostructures, correlated insulating states have been shown to emerge for different fractional charge fillings of moiré superlattice sites, enabling investigation of quantum many-body states \cite{tang2020simulation, regan2020mott, xu2020correlated, huang2021correlated, jin2021stripe, li2021imaging} with the potential for optical measurement and coherent control of strongly-correlated phases \cite{jin2019observation, seyler2019signatures, tran2019evidence, alexeev2019resonantly, baek2020highly, wang2020correlated}. Twisted \textit{homo}structures not only allow for controlled tuning of the underlying moiré superlattice potential but also enhanced formation of hybridized minibands due to the absence of lattice and energy mismatch in the constitutive monolayers \cite{Plo20}. Until now, few studies have appeared on the angle-dependent optical and electronic properties of twisted $\mathrm{MoSe}_2$ homobilayers including photoluminescence (PL) experiments exhibiting moiré-trapped trions for twist angles close to \ang{0} and \ang{60} \cite{Mar21}; and a static electric dipole moment characterization of different excitonic species on a single $\mathrm{MoSe}_2$ bilayer with a fixed twist angle of \ang{0} \cite{Sun20}. Recent reports on $\mathrm{WS}_2$ \cite{Yan19} and $\mathrm{MoS}_2$ \cite{Van14,Liu14} showed that the \textbf{K}-\textbf{K} exciton transition \cite{Yan19,Van14,Liu14} is insensitive to twist angle. 

We combine optical spectroscopy on hBN-encapsulated $\mathrm{MoSe}_2$ homobilayers with theory to obtain new information about the twist angle dependent optoelectronic response. Both, direct \textbf{K}-\textbf{K} and indirect \boldsymbol{$\Gamma$}-\textbf{K} excitons, exhibit an abrupt decrease of the emission energy in the vicinity of \ang{0} or \ang{60}. This rapid change of exciton energy is accompanied by the appearance of indirect exciton PL below \ang{8} and above \ang{54} and vanishing direct exciton PL. Our theoretical predictions based on an ab-initio-based continuum model are in excellent agreement with experiment, showing that the transition between different regimes of emission energy can be understood in terms of increasing exciton localization in moiré sites at small twist angles. Furthermore, our model allows us to characterize the observed exciton resonances, providing insight into their layer composition and resulting binding energies.

\begin{figure*}[t]
\includegraphics{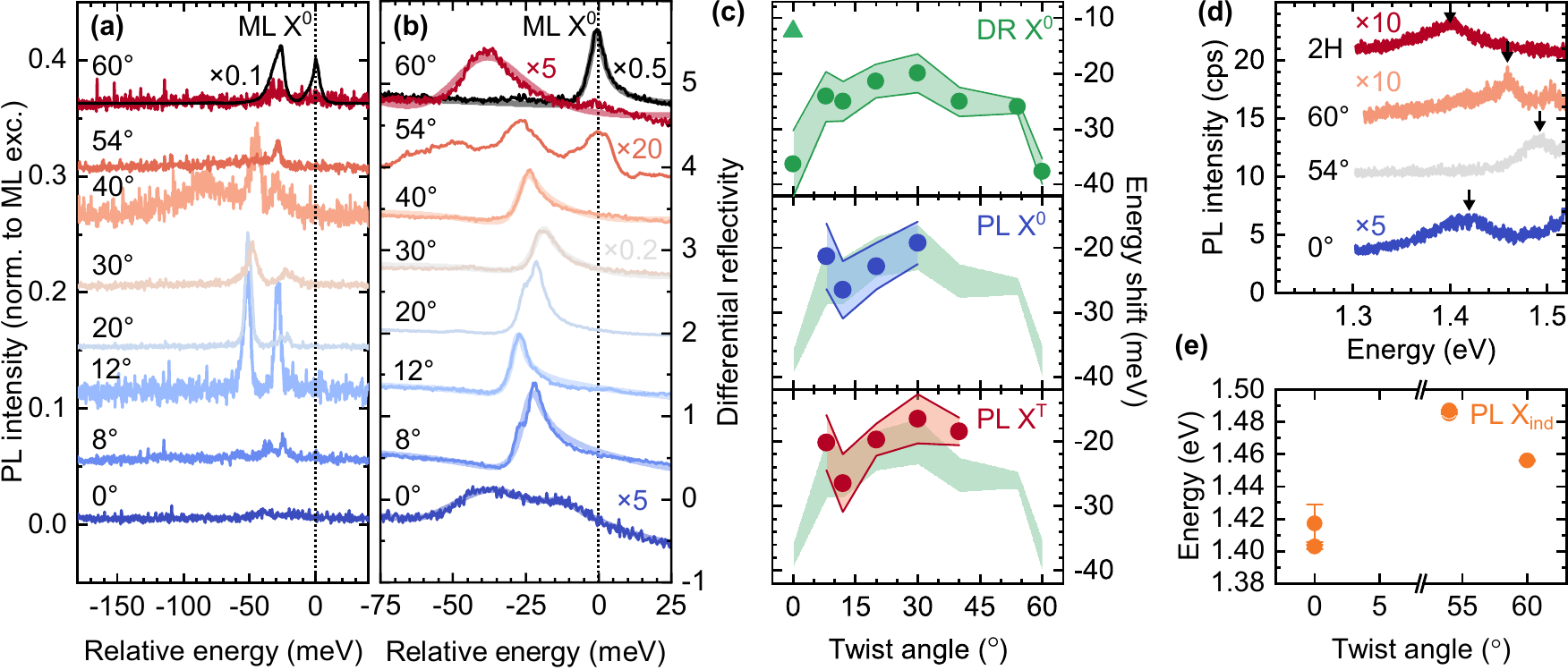}
\caption{\label{f1}
Photoluminescence \textbf{(a)} and reflectivity \textbf{(b)} spectra for $\mathrm{MoSe_{2}}$ homobilayers with different twist angles. The energies are given as the detuning with respect to the neutral exciton in the monolayer region of each sample, labeled by dotted lines. Black spectra in both panels show an exemplary monolayer signal. \textbf{(c)} Energy shift between monolayer and bilayer regions for excitons and trions extracted from the data presented in panels (a) and (b). Green (blue) data points represent the neutral exciton shift extracted from reflectivity (PL) measurements. Red data points represent trion shifts extracted from PL measurements. Error bands for the extracted spectral positions are respectively color coded with the reflectivity measurements error band being extended across all panels for comparison. Negative shifts correspond to an energy reduction in the bilayer region with respect to the monolayer region. Indirect exciton (X$_\mathrm{ind}$) PL spectra \textbf{(d)} and extracted energies \textbf{(e)} as a function of twist angle of the MoSe$_2$ homobilayers.
}
\end{figure*}

Samples were prepared using a tear-and-stack technique \cite{Kim16} combined with a modified version of the hot-pick-up method \cite{Wan13, Piz16} to assemble twisted MoSe$_2$ homobilayers with  relative stacking angle $\Delta\theta$ (see the Supplemental Material (SM)). 
Fig.~\ref{f1}(a) presents optical PL emission from $\mathrm{MoSe_{2}}$ twisted bilayers in the range $\SI{0}{\degree} \leq\Delta\theta \leq \SI{60}{\degree}$, probed at \SI{10}{\kelvin} using \SI{500}{nW} of continuous-wave \SI{532}{nm} excitation laser focused onto a diffraction-limited spot (100$\times$ objective, NA=0.7). 
PL spectra in Fig.~\ref{f1}(a) are plotted as a function of neutral \textbf{K}-\textbf{K} exciton ($\mathrm{X}^{0}$) detuning between homo- and monolayer regions and normalized with respect to $\mathrm{X}^{0}$ intensity measured in the respective monolayers. We observe that $\mathrm{X}^{0}$ and trion ($\mathrm{X}^{\mathrm{T}}$) energies are redshifted relative to the monolayer $\mathrm{X}^{0}$.
One striking feature of the data presented in Fig.~\ref{f1}(a) is the complete absence of $\mathrm{X}^{0}$ and $\mathrm{X}^{\mathrm{T}}$ emission for bilayers having twist angles in the ranges \ang{0} to \ang{8} and \ang{54} to \ang{60}.

\begin{figure*}[t]
\includegraphics{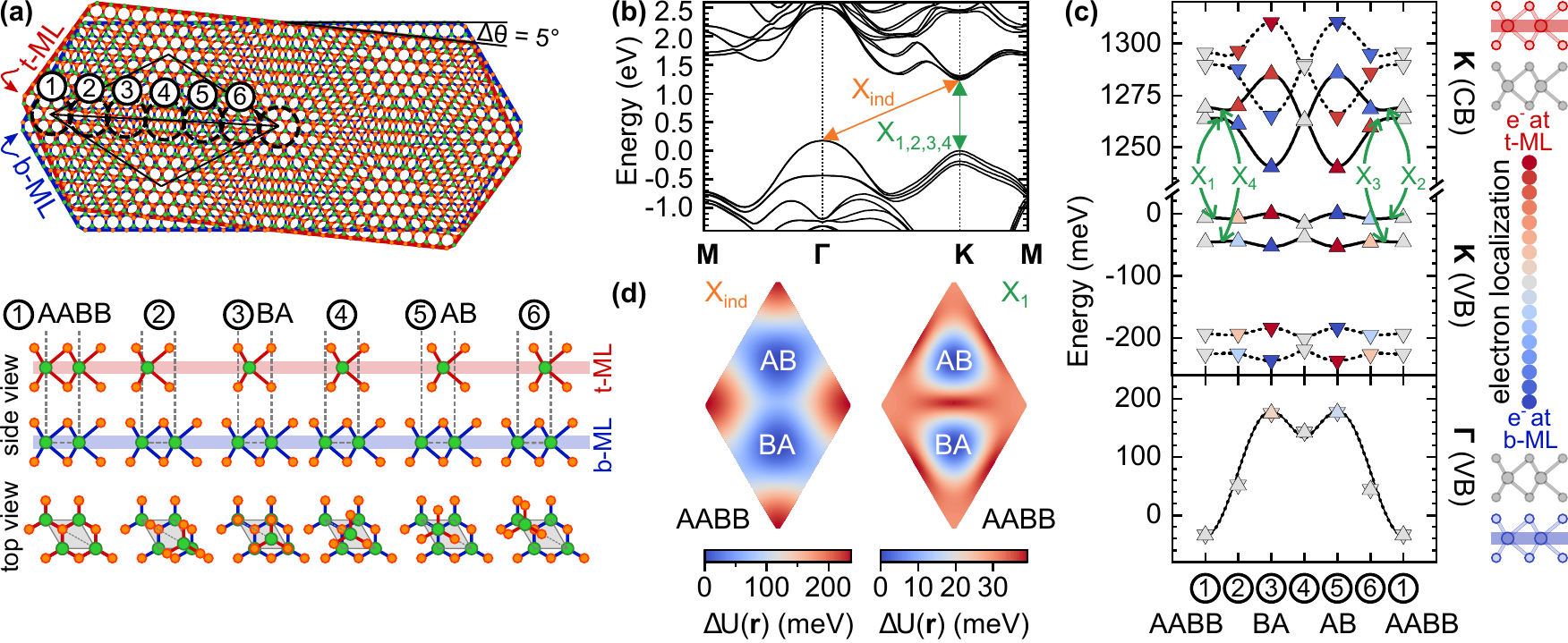}
\caption{\label{f2}
\textbf{(a)} Top panel: Moiré pattern formed by superposition of two hexagonal lattices twisted by $\Delta\theta=\ang{5}$ showing alternating regions of AABB, AB and BA stacking order. Bottom panel: Side view of the local stacking configurations marked in the left panel. \textbf{(b)} DFT band structure for AB stacking including spin-orbit coupling. The green (orange) arrow indicates the studied direct (indirect) transitions. \textbf{(c)} Variation of the conduction band at the \textbf{K}-point as well as the valence bands at the \textbf{K}- and \boldsymbol{$\Gamma$}-point as a function of interlayer translation. Up (down) triangles correspond to spin up (down) bands obtained from individual DFT calculations for stacking configurations defined in (a). The color encodes the layer contribution to Bloch states. The solid (dashed) lines show the result of our fit model for the variation of spin up (down) bands. \textbf{(d)} Moiré potentials $\Delta U(\textbf{r})$ for the indirect exciton $\mathrm{X_{ind}}$ and the direct 
exciton $\mathrm{X_1}$ shown relative to the potential minimum within the moiré unit cell.
}
\end{figure*}

Fig.~\ref{f1}(b) presents differential reflectivity measurements for the same series of samples discussed in Fig.~\ref{f1}(a). We note that the reflectivity data reveals clear signatures of the $\mathrm{X}^{0}$ transition from samples having small twist angles (\ang{0} to \ang{8} and \ang{54} to \ang{60}) which could not be observed in PL experiments due to the low quantum yield of these transitions \cite{frisenda2017micro,Eva71, Ane80}.
To analyze the reflectivity spectra, we modeled the refractive index of the MoSe$_2$ monolayer as a single Lorentz oscillator, except in the case of \ang{0} where we used two oscillators, and calculated the differential reflectivity using the transfer-matrix method (see the SM). 
The simulated reflectivity curves are shown as diffuse color lines in Fig.~\ref{f1}(b) with good agreement between experiment and theory. The reflectivity of \ang{0}, \ang{8} and \ang{20} show a double-peak structure predicted by our theoretical model steaming from different hybridized \textbf{K}-\textbf{K} excitons, as we will see below. 
We evaluated the average energies of $\mathrm{X}^{0}$ and $\mathrm{X}^{\mathrm{T}}$, as obtained from PL and reflectivity measurements. In Fig.~\ref{f1}(c), we present the energy shift as a function of the stacking angle, relative to the energy of the corresponding excitonic transition ($\mathrm{X}^{0}$ or $\mathrm{X}^{\mathrm{T}}$) in the monolayer. Green data points within the green error band in the uppermost panel in Fig.~\ref{f1}(c) denote relative $\mathrm{X}^{\mathrm{0}}$ energies measured via differential reflectivity. Blue and red data points in the middle and bottommost panel show the energy shift of $\mathrm{X}^{0}$ and $\mathrm{X}^{\mathrm{T}}$ respectively, determined from PL measurements. At large twist angles (away from \ang{0} and \ang{60}), $\mathrm{X}^{0}$ and $\mathrm{X}^{\mathrm{T}}$ exhibit similar behavior: Both are redshifted by $\sim$\SI{20}{meV} with respect to the monolayer emission with only a minor twist angle dependence. 
The observed energetic reduction for $\mathrm{X}^{0}$ and  $\mathrm{X}^{\mathrm{T}}$ transitions in the homobilayer region can be explained considering two fundamentally different effects: (i) hybridization of electronic states between layers \cite{Li07,Spl10} and (ii) static dielectric screening induced by hBN encapsulation and the mutual proximity of the second MoSe$_2$ monolayer \cite{Uge14, Raj17, Flo18, Cho18}. 
In fact, $\mathrm{X}^{0}$ energy has a strong sensitivity to its dielectric environment enhanced by the bilayers reduced dimensionality and dielectric constant mismatch between hBN and MoSe$_2$ \cite{segura2018natural}. 
Consequently, for stacking angles around $\Delta\theta \simeq$ \ang{30}, static screening is the dominant effect inducing a constant redshift in the $\mathrm{X}^{0}$ and $\mathrm{X}^{\mathrm{T}}$ energies \cite{Cho18,Lin14, Sti16, Flo18}. 
Remarkably, as twist angles of \ang{0} or \ang{60} are approached the redshift increases to $\sim\SI{40}{\milli\eV}$. In this case, hybridization effects between the layers are dominant and twist-angle dependent.
Figs.~\ref{f1}(d) and (e) present measured PL spectra and energies for indirect \boldsymbol{$\Gamma$}-\textbf{K} excitons (X$_\mathrm{ind}$), appearing $\SI{200}{\milli\eV}$ below the observed $\mathrm{X}^{0}$ transition. We observe a strong energy dependence as a function of stacking angle, similarly as the one found for $\mathrm{X}^{0}$ and $\mathrm{X}^{\mathrm{T}}$.
For twist angles in the ranges \ang{0} to \ang{8} and \ang{54} to \ang{60}, the decrease in the intensity of the $\mathrm{X}^{0}$ PL emission is accompanied by an increase on PL emission into these X$_\mathrm{ind}$ momentum indirect states at lower energies \cite{Sun20}. 
As we now continue to show, the experimentally-observed $\mathrm{X}^{0}$ and $\mathrm{X}^{\mathrm{T}}$ energetic shifts and X$_\mathrm{ind}$ PL emission increase for low stacking angles originate from moiré physics, a behaviour predicted by our low-energy continuum model.

For small stacking angles, the moiré pattern can be seen as a smooth variation of the stacking configurations known from untwisted bilayers (see Fig.~\ref{f2}(a)). 
In this case, local stacking configurations found on a path across the supercell can be approximated by translations of the top layer with respect to the bottom layer, starting from the AABB configuration and passing through stacking configurations AB and BA. We use the generic term R-stacking for all configurations of this type. Likewise, for the near-\ang{60} case (H-stacking), we introduce a translation through AA, ABBA and BB stacking.  
Contrary to rotations, translations do not change the periodicity of the bilayer so that density functional theory (DFT) calculations are readily performed for the different stacking configurations. For both, R- and H-stacking, we choose six translations corresponding to six equidistant points within one supercell.
We use a fixed lattice constant of $a=\SI{3.336}{\angstrom}$, which accounts for \SI{1.4}{\percent} of strain compared to a naturally-grown homobilayer. Similar values have already been proposed for MoSe$_2$ homobilayers \cite{Sun20} and are consistent with strain maxima of up to \SI{2.1}{\percent} recently calculated for twisted homobilayers \cite{linderalv2022moire} (see the SM). Fig.~\ref{f2}(b) introduces the lowest bright \textbf{K}-\textbf{K} transitions $\mathrm{X_1}$, $\mathrm{X_2}$, $\mathrm{X_3}$, and $\mathrm{X_4}$ and the lowest indirect \boldsymbol{$\Gamma$}-\textbf{K} exciton ($\mathrm{X_{ind}}$) calculated for AB stacking including spin-orbit coupling.
The results for the twist-dependent \boldsymbol{$\Gamma$}- and \textbf{K}-point energies are presented in Fig.~\ref{f2}(c). 
We find that electron localization at the \textbf{K}-point depends strongly on the stacking configuration. While for high-symmetry configurations AABB we find perfect electronic delocalization over both layers, electrons are well localized in one of the two layers for the most stable configurations AB and BA. However, we note that this phenomenon is exclusive for R-stacking and limited by spin-layer polarization in case of H-stacking. For the latter, well-localized states are found for all configurations (see the SM). 
Compared to the \textbf{K}-point, the \boldsymbol{$\Gamma$}-point energy shows a much stronger stacking dependence. This modulation is directly related to the varying interlayer distance for different stacking configurations, which has particularly high influence on the \boldsymbol{$\Gamma$}-point due to strong interlayer hopping. Moreover, interlayer hopping leads to a pronounced spread of the electronic states over both layers as indicated by the constant light gray color of the corresponding data points in Fig.~\ref{f2}(c). Even though there are several indirect excitons involving higher conduction bands at the \textbf{K}-point, the variation of the \boldsymbol{$\Gamma$}-point dominates the corresponding moiré potential, which in turn leads to negligible variations between the different indirect excitons. 

To calculate the individual moiré potentials, we model the band variations of Fig.~\ref{f2}(c) by symmetry considerations using a similar approach to Refs. \cite{wang2017interlayer, hagel2021exciton} (see the SM). Thereby, we have direct access to a set of periodic band gap variations $U(\pmb{r})$, each defined by two high-symmetry points and corresponding bands. Excitons inside the moiré superlattice experience this energetic modulation effectively as the moiré potential, which is visualized for the exciton species $\mathrm{X_{ind}}$ and $\mathrm{X_1}$ in Fig.~\ref{f2}(d). In this picture, the Hamiltonian for the center-of-mass (COM) motion of the exciton is given by \cite{wu2018theory}
\begin{equation}
\label{eq:MacDonald}
   \hat{H}_0 = E^0_\mathrm{gap}+\frac{\hbar^2 \pmb{\hat{Q}}^2}{2M}+U(\pmb{\hat{r}})\,,
\end{equation}
where $E^0_\mathrm{gap}$ is the local band gap corresponding to $\pmb{r}=0$ and $M$ is the total effective mass of electron and hole combined. $\pmb{Q}$ and $\pmb{r}$ are the COM wavevector and position vector, respectively. In our model, the twist angle only enters via a scaling factor inside the moiré potential $U(\pmb{r})$ acting on the argument $\pmb{r}$. This dependence results from describing every point inside the supercell by its local and untwisted band structure configuration. Thus, for the near-\ang{0} case, we assume the moiré supercell to scale with $1/\sin(\theta/2)$ and for the near-\ang{60} case with $1/\sin((\frac{\pi}{3}-\theta)/2)$ \cite{moon2013optical}. Outside the moiré regime, approximately between \ang{10} and \ang{50}, our calculations can be interpreted as an extrapolation of the moiré case.
Even though this approximation is not strictly valid for large twist angles in the vicinity of \SI{30}{\degree}, our calculations are useful to identify differences between R- and H-stacking. The exciton COM motion is accompanied by a relative motion of electron and hole described by the Wannier equation, which accounts for the screened electron-hole Coulomb potential. The latter depends on the actual charge distribution in stacking direction and is therefore calculated separately for the case of intra- and interlayer excitons.\cite{wang2019optical} Then, the admixture of intra- and interlayer character is taken into account according to the layer contributions of valence and conduction band given by the colormap in Fig.~\ref{f2}(c) (see the SM).

\begin{figure}
\includegraphics{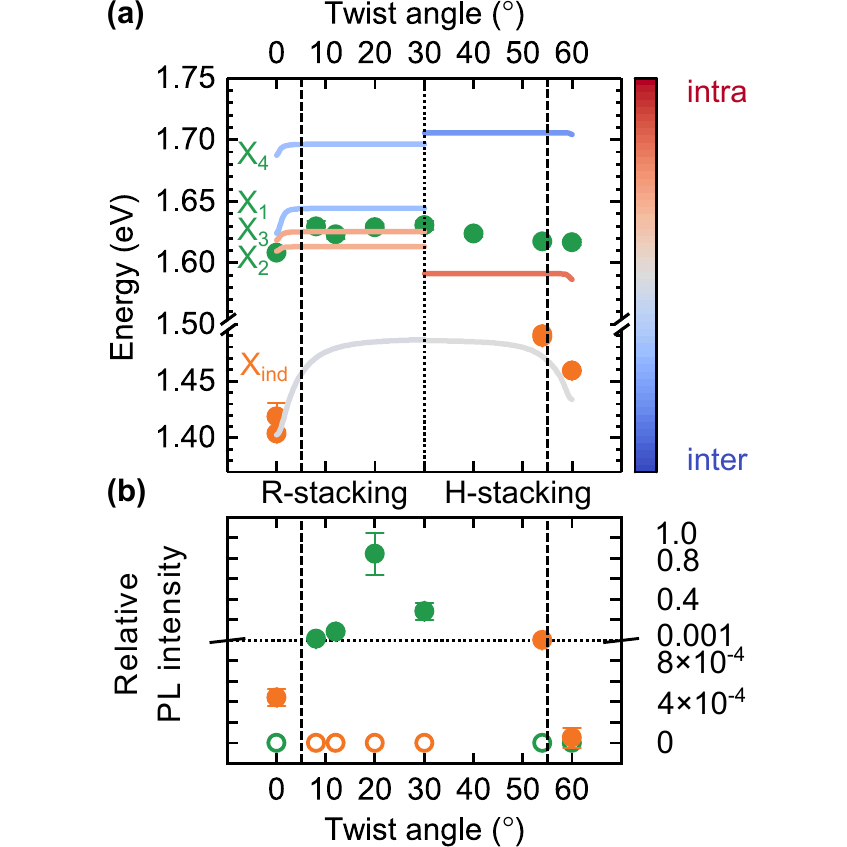}
\caption{\label{f3}
\textbf{(a)} Angle-dependent exciton energies. Solid lines show the result of our continuum model based on DFT calculations. Each line corresponds to an individual exciton species, while the color encodes the layer character.
The intermediate case (gray color) indicates delocalization of electron and/or hole over both layers. The green and orange data points are taken from Fig.~\ref{f1}(c) (DR) and Fig.~\ref{f1}(e) (PL), respectively.
\textbf{(b)} Integrated PL intensity for $\mathrm{X}^{0}$ and $\mathrm{X_{ind}}$ relative to the corresponding monolayer intensity as a function of twist angle.
}
\end{figure}

The main findings of our work are presented in Fig.~\ref{f3}. Panel (a) compares the measured ($\mathrm{X}^{0}$, $\mathrm{X_{ind}}$, data points) and calculated  ($\mathrm{X_{1,2,3,4}}$, $\mathrm{X_{ind}}$, solid lines) transition energies as a function of twist angle. The theoretical curves are uniformly shifted by $\SI{554}{\milli\eV}$ to match the experimental data point of the indirect exciton for small twist angles, which is in good agreement with expected GW corrections for $\mathrm{MoSe_{2}}$ homobilayers \cite{he2014stacking} (see the SM). The energy discontinuity that can be seen in Fig.~\ref{f3}(a) around \ang{30} arises from the delicate distinction between intra- and interlayer excitons, which is clearer for H-stacking due to the aforementioned spin-layer polarization. This effect leads to a strong intralayer exciton binding energy and thereby results in a slight asymmetry when comparing the energies approaching from R- and H-stacking configurations at \ang{30}.
Since the layer contributions to each exciton directly influence the binding energies, we visualized them separately via the line colors. We find that for both, R- and H-stacking, the two lower direct excitons $\mathrm{X}_{2}$ and $\mathrm{X}_{3}$ have mostly intralayer character, while interlayer excitons $\mathrm{X}_{1}$ and $\mathrm{X}_{4}$ have higher energies due to a reduced Coulomb interaction. In case of R-stacking, intralayer excitons $\mathrm{X}_{2}$ and $\mathrm{X}_{3}$ are split by \SI{12}{\milli\eV}, closely followed by the lowest interlayer exciton $\mathrm{X}_{1}$. Thereby, $\mathrm{X}_{1}$, $\mathrm{X}_{2}$ and $\mathrm{X}_{3}$ are in agreement with the experimental findings of Ref. \cite{Sun20} for angles near \ang{0}. For H-stacking, inversion symmetry leads to double degeneracy of intra- and interlayer excitons. The model explains our experimental findings, and we could assign the measured $\mathrm{X}^{0}$ to $\mathrm{X}_{2}$. However, the reflectivity data presented in Fig.~\ref{f1}(b) for \ang{0}, \ang{8} and \ang{20} samples hints at a double peak structure of $\mathrm{X}^{0}$ that might originate from the R-type $\mathrm{X}_{2}$ and $\mathrm{X}_{3}$ intralayer excitons. In that case, the measured $\mathrm{X}^{0}$ could arise from an hybridized mixture of these two species \cite{ruiz2019interlayer}.
For small stacking angles, our model predicts a shift of \SI{17}{\milli\eV} spanned by both exciton species, a value that matches very well the experimental shift obtained for the $\mathrm{X}^{0}$ resonance of \SI{18}{\milli\eV}.
The small discrepancy between theory and experiment may be due to lattice reconstruction effects \cite{weston2020atomic} that are not taken into account in our models. 

As depicted in Fig.~\ref{f3}(a), the indirect \boldsymbol{$\Gamma$}-\textbf{K} transition $\mathrm{X_{ind}}$ possesses a very strong angle dependence for small stacking angles, resulting from the interaction with the deep moiré potential shown in the left panel in Fig.~\ref{f2}(d). Towards \ang{30} the moiré supercell becomes incommensurately small and the interlayer distance in the homobilayer increases. In this regime, the kinetic energy dominates the COM Hamiltonian (\ref{eq:MacDonald}), resulting in a quasi-homogeneous spreading of the wave function across the whole superlattice and therefore a weak angular dependence. The critical angle for delocalization is smaller for direct excitons ($\approx\SI{3}{\degree}$) and comparable to reported findings in twisted heterobilayers \cite{brem2020tunable}.

Fig.~\ref{f3}(b) presents the integrated experimental PL intensity for $\mathrm{X}^{0}$ and $\mathrm{X_{ind}}$ as a function of twist angle, normalized to the intensity of monolayer $\mathrm{X}^{0}$ for each sample.
Interestingly, for angles in the range from \ang{0} to \ang{5} and \ang{55} to \ang{60}, we observe a PL intensity downconversion from direct \textbf{K}-\textbf{K} excitons to indirect \boldsymbol{$\Gamma$}-\textbf{K}. For these small stacking angles, we calculate an increase of the splitting between $\mathrm{X}^{0}$ and $\mathrm{X_{ind}}$  from approximately \SI{150}{\milli\eV} to \SI{200}{\milli\eV}, favouring a nonradiative decay of $\mathrm{X}^{0}$ to the lowest $\mathrm{X_{ind}}$ excitons and thus inducing the shift of the PL emission intensity.

In conclusion, we presented a systematic investigation of optical properties of hBN-encapsulated $\mathrm{MoSe}_2$ homobilayers as a function of the interlayer twist angle. To explain our experimental observations, we performed moiré superlattice exciton calculations using an ab-initio-based continuum model. 
Entirely new information was obtained on the optical and electronic properties of different excitonic species in MoSe$_2$ homobilayers, providing insight into layer composition and relevant stacking configuration of different moiré excitonic species. Controlled engineering of these properties could lead towards the realization of on-chip quantum simulators based on periodic excitonic arrays.
Novel strongly-correlated excitonic states could become readily available for exploration, such as exciton liquids and two-dimensional condensates.

\begin{acknowledgments}
We gratefully acknowledge financial support from the German Science Foundation (DFG) via projects FI 947/8-1, DI 2013/5-1 and SPP-2244. Moreover, research was partly funded via the Excellence Initiative in the framework of the projects e-conversion (EXS 2089) and the Munich Center for Quantum Science and Technology (MCQST EXS 2111). J.J.F. thanks the DFG for financial support via the projects INST95-1496-1 and INST95-1642. M.K. gratefully acknowledges funding from the International Max Planck Research School for Quantum Science and Technology (IMPRS-QST). M.M.P. acknowledges TUM International Graduate School of Science and Engineering (IGSSE). 
V.V. and M.F. acknowledge support by the Alexander von Humboldt foundation. The authors would also like to thank Prof. M. Kira for valuable discussions.
\end{acknowledgments}

\bibliography{ref}

\end{document}